\documentclass[letterpaper]{article} 
\usepackage{aaai24}  
\usepackage{times}  
\usepackage{helvet}  
\usepackage{courier}  
\usepackage[hyphens]{url}  
\usepackage{graphicx} 
\usepackage{natbib}  
\usepackage{caption} 
\usepackage{algorithm}
\usepackage{algorithmic}

\usepackage{newfloat}
\usepackage{listings}
\usepackage{amsfonts}
\usepackage{amsmath}
\usepackage{booktabs}
\usepackage{multirow}
\nocopyright
\DeclareCaptionStyle{ruled}{labelfont=normalfont,labelsep=colon,strut=off} 
\floatstyle{ruled}
\newfloat{listing}{tb}{lst}{}
\floatname{listing}{Listing}
\title{UIPC-MF: User-Item Prototype Connection Matrix Factorization for Explainable Collaborative Filtering}
\author{
    Lei Pan\textsuperscript{\rm *},
    Von-Wun Soo\textsuperscript{\rm *}\textsuperscript{\rm †}
}
\affiliations{
    \textsuperscript{\rm *}National Tsing Hua University\\
    \textsuperscript{\rm †}Chang Gung University\\

    parrotlet@gapp.nthu.edu.tw, soo@cgu.edu.tw
}

\usepackage{bibentry}

\begin{document}

\maketitle

\begin{abstract}
Recommending items to potentially interested users has been an important commercial task that faces two main challenges: accuracy and explainability. While most collaborative filtering models rely on statistical computations on a large scale of interaction data between users and items and can achieve high performance, they often lack clear explanatory power. We propose UIPC-MF, a prototype-based matrix factorization method for explainable collaborative filtering recommendations. In UIPC-MF, both users and items are associated with sets of prototypes, capturing general collaborative attributes. To enhance explainability, UIPC-MF learns connection weights that reflect the associative relations between user and item prototypes for recommendations. UIPC-MF outperforms other prototype-based  baseline methods in terms of Hit Ratio and Normalized Discounted Cumulative Gain on three datasets, while also providing better transparency.
\end{abstract}

\section{Introduction}

Recommending items to potential interested users has been an important commercial task that faces with two main challenges the accuracy and explainability. The well-known collaborative filtering methods that typically used to recommend items for a particular users rely on statistical computing on a large scale of interaction data between users and items and can achieve high performance. However, the sophisticated hidden underlying computing has made the methods vague in explaining the rationales of its recommendation. Therefore, making recommendation explainable and interpretable has become a performance metric for an recommender system in addition to its accuracy of recommendation prediction.

Prototypes are representative examples that involve the learning of latent representations from the data space that were widely adopted in interpretable machine learning \cite{molnar22, Zhang21} and had extensive applications in computer vision \cite{Li18, Chen19, Nauta21, Keswani22}. These approaches leverage prototypes to extract significant patterns from large images and determine similarity scores between a specific photo and learned prototypes in order to facilitate effective identification. The features of the prototypes can serve as the interpretable basis for a group of typical objects.

However, selecting or identifying the proper prototypes can be critical in both accuracy and interpretability in recommendation. We propose User-Item Prototypes Connections Matrix Factorization (UIPC-MF), a novel prototype-based collaborative filtering algorithm. UIPC-MF learns two sets of prototypes for both users and items from large interaction data, with each prototype capturing a common attribute derived from collective wisdom. A predicted logit score is computed through a linear combination of user similarities between a user representation and each user prototype, as well as item similarities between an item representation and each item prototype. The score computation is conducted based on the connection weights among the user prototypes and item prototypes. This novel approach allows transparent understanding of how the predicted logit score is composed, offering more insights than other prototype-based methods. The transparency makes the recommender systems explainable as it enables the model designers to understand the recommender system's underlying operations and effects, and thus can provide users with rationales of a recommendation \cite{Tintarev15, Chen22, Balog20}.

In optimization of the UIPC-MF model, due to the large training parameter space that may lead to local optimality of the model and the training process can be also biased to certain states that lead to biased interpretation. It is important to setup regularization term in the objective loss function in order to avoid the biased results.

The objectives of the paper focus on the issues around learning the models of the prototype representation connection mapping between the user space and item space in order to make accurate and transparent recommendations given user interaction history data. They can be formulated as the following three research questions:

\begin{itemize}
\item \textbf{RQ1.}How does the performance of our recommendation model compare with the state-of-the-art baseline model?
\item \textbf{RQ2.}How can we understand the underlying inference of our model and to generate explanation rationales for the recommendations?
\item \textbf{RQ3.} How can we setup the loss objective functions as well as proper regularization term in avoid of the potential learning bias of the model in the optimization of the parameters of the model against the given training data?
    
\end{itemize}

\section{Related Work}
\subsection{Collaborative filtering}

The collaborative filtering based models tracking the user's history or feedback (e.g. ratings or clicks) to evaluate user's preference and relations to items. The difference of techniques among different collaborative filtering based models is the ways of calculating the item-user association scores based on the user-item interaction history. When the user-item pair is fed into the kind of models, it returns an association score in which the higher score indicates higher preference of the user to the item. To acquire such an user-item association relation, the most popular technique is the matrix factorization (MF) methods  \cite{Koren09, Rendle09} which the user-item interactions construct an original user-item interaction matrix.

Recently, deep learning models have been widely used in recommendation systems due to their high performance in learning implicit representations. These systems utilize deep learning models to learn user/item representations and interaction functions to estimate the user-item scores. For example, in \cite{He17} they used multi-layer perceptrons to learn the interaction function rather than using the traditional inner product. In \cite{Liang18}, they utilized a variational autoencoder for collaborative filtering and learned a multinomial likelihood distribution for the user-item interaction. In \cite{He20}, they applied graph convolution networks to collaborative filtering. Their results outperformed traditional MF-based methods.

However, these algorithms, regardless of their association accuracy, are based on sophisticated statistical correlation calculation that is hard to understand the underlying inference mechanisms for the final recommendation and thus become explanation opague.
In order to make the correlation calculation more explicit and interpretable, we propose prototype-based collaborative filtering method to be discussed in the following section.

\subsection{Explainable Recommender Models}

Transparency of a recommender system is important to allow users or developers to understand the rationales of underlying operations or inferences of the recommender system \cite{Tintarev15, Chen22, Balog20}. An explainable recommender system refers to generating explanations for the predicted outcomes of the recommender system. An explainable recommender system can provide users and developers with a reason for why a particular recommendation was made. An explainable system may not necessarily be totally transparent.

The explainable recommender models can be categorized into two types based on their levels of transparency \cite{Zhang20}. The first is the model-intrinsic approach \cite{molnar22}, where the recommendation mechanism is transparent by design and can be utilized to generate explanations. For example, Chen et al. \cite{Chen18} proposed the Attention-driven Factor Model (AFM), which incorporates a learnable attention mechanism for user to consider item features in the recommendation model. Another example of a model-intrinsic approach is the prototype-based models such as UIPC-MF, which learn representative examples within the model's internal workings. The latter is the model-agnostic approach, also known as a post-hoc approach. \cite{molnar22}  In this approach, the recommendation model is a blackbox by itself that requires to implement another explanation model to explain the recommendations made by the recommendation model \cite{Zhang20}. For instance, Wang et al. \cite{Wang18} proposed a reinforcement learning framework for explainable recommendation. They design a black box agent to generate the predicted recommendation rating score, and two explainable agents that receive the state from the black box agent and environment to generate the explanation and predicted rating. Finally, the explainable component can learn to explain and predict. 

\subsection{The Prototype-based Collaborative Filtering}\label{sec:related_work_proto}

The concept of using prototypes to learn representative vectors for users and items has recently been employed in Anchor-based Collaborative Filtering (ACF) \cite{Barkan21} and Prototype-based Matrix Factorization (ProtoMF) \cite{Melchiorre22}. ACF learns shared vectors known as anchors, and users and items are represented as convex combinations of these anchor vectors. The predicted output in ACF is obtained through the dot product between the user representation and the item representation. However, this output logit does not directly reflect a linear relationship with the anchors that can limit the interpretability.

In ProtoMF, prototypes are learned separately for users and items. The predicted logit in ProtoMF is computed by summing the dot products between user similarities from each user prototype and item representation with linear transformation weights, as well as the dot products between item similarities from each item prototype and user representation with linear transformation weights.  Although the logit output of ProtoMF can be decomposed into a linear combination of user prototype similarities and item prototype similarities, the lack of connections between user prototypes and item prototypes makes it unable to directly associate the relationship between them as well as distinguish user's preference over each item prototype.

In summary, these methods learn representative prototype vectors from user vectors and item vectors, where the user vectors and item vectors have direct relationships with the prototype vectors. However, there is a lack of clear and direct relationships between the user prototype vectors and the item prototype vectors, which limits transparency. To address this problem, UIPC-MF introduces weights between user prototype vectors and item prototype vectors. The predicted scores are then calculated as a linear combination of user prototype similarities and item prototype similarities, resulting in improved transparency.

\section{Methodology}

In this section, we first provide a detailed description on on our proposed model, User-Item Prototypes Connections Matrix Factoriza-tion (UIPC-MF), and then explain the objective function. UIPC-MF is specifically designed for the top-N recommendation based on implicit feedback, such as user's listening events and click history, rather than explicit feedback like user ratings. The goal of top-N recommendation is to provide each user with a small set of N items from a large pool of available items \cite{Cremonesi10, Zolaktaf18}. In top-N recommendations, a higher logit of the predicted output indicates a greater likelihood of user interaction with the recommendation.

\subsection{User-Item Prototypes Connections Matrix Factorization (UIPC-MF)}
The aim of UIPC-MF is to identify collaborative patterns among users and items based on interaction data in terms of prototype vectors. For example, from the perspective of user prototypes, an individual who likes metal music may have an embedding vector that is closer to a user prototype with an attribute being a metal lover. Similarly, from the perspective of item prototypes, the embedding vector of a metal music piece will be closer to an item prototype with an attribute belonging to the metal genre.

Firstly, we define $\mathcal{U} \text{ = } \{{u_{i}}\in\mathbb{R}^{d}\}_{i=1}^{N}$ and $\mathcal{T} \text{ = } \{{t_{j}}\in\mathbb{R}^{d}\}_{j=1}^{M}$ as the set of N users and M items, respectively. Let ${\mathcal{P}^{u}} \text{ = } \{{p_{i}^{u}}\in\mathbb{R}^{d}\}_{i=1}^{L^{u}}$ be the set of ${L^{u}}$ learnable user prototypes, where ${L^{u} \ll N}$, and ${\mathcal{P}^{t}} \text{ = } \{{p_{i}^{t}}\in\mathbb{R}^{d}\}_{i=1}^{L^{t}}$  be the set of ${L^{t}}$ learnable item prototypes, where ${L^{t} \ll M}$. ${L^{u}}$ and ${L^{t}}$ are hyperparameters. For the implicit interaction data, $\mathcal{I}\text{ = }\{(u_{i},t_{j})\}$, where $(u_{i},t_{j})$ represents that the user ${u_{i}}$ has interacted with the item ${t_{j}}$. For simplicity, we omit the indices of users or items to represent any user or item.

To obtain the collaborative patterns of user $u$, we adopt the method proposed in \cite{Melchiorre22}, which calculates the user-prototype similarity vector $u^{*}$  based on the user vector and each user prototype vector. UIPC-MF first calculates the similarity scores between a user $u$ and each user prototype $p^{u}$, resulting in ${u^{*}\in\mathbb{R}^{L^{u}}}$, a user-prototype similarity vector. The similarity function used is the shifted cosine similarity, which has a range of 0 to 2. Similarly, UIPC-MF calculates ${t^{*}\in\mathbb{R}^{L^{t}}}$, an item-prototype similarity vector, using item $t$ and each item prototype vector $p^{t}$ in the similar manner as $u^{*}$.  The index $i$ in $u^{*}$ represents the similarity score between a user embedding $u$ and the i-th user prototype vector $p_{i}^{u}$, allowing $u^{*}$ to capture all the scores of the user's collaborative patterns. The same concept applies to ${t^{*}}$. The formulas for calculating ${u^{*}}$ and ${t^{*}}$ are shown in equation \ref{eq:3.1}.

\begin{equation} \label{eq:3.1}
\begin{split}
{u^{*}}=\begin{bmatrix}
 sim(u,p_{1}^{u})\\
 ...\\
 sim(u,p_{L^{u}}^{u})
\end{bmatrix}
\in\mathbb{R}^{L^{u}},\; {t^{*}}=\begin{bmatrix}
 sim(t,p_{1}^{t})\\
 ...\\
 sim(t,p_{L^{t}}^{t})
\end{bmatrix}
\in\mathbb{R}^{L^{t}}\\, \;where \; sim(a,b) = 1+\frac{a^{T}b}{\left\| a\right\|\cdot\left\| b\right\|}
\end{split}
\end{equation}

\begin{figure}[H]
  \centering
  \includegraphics[width=\linewidth]{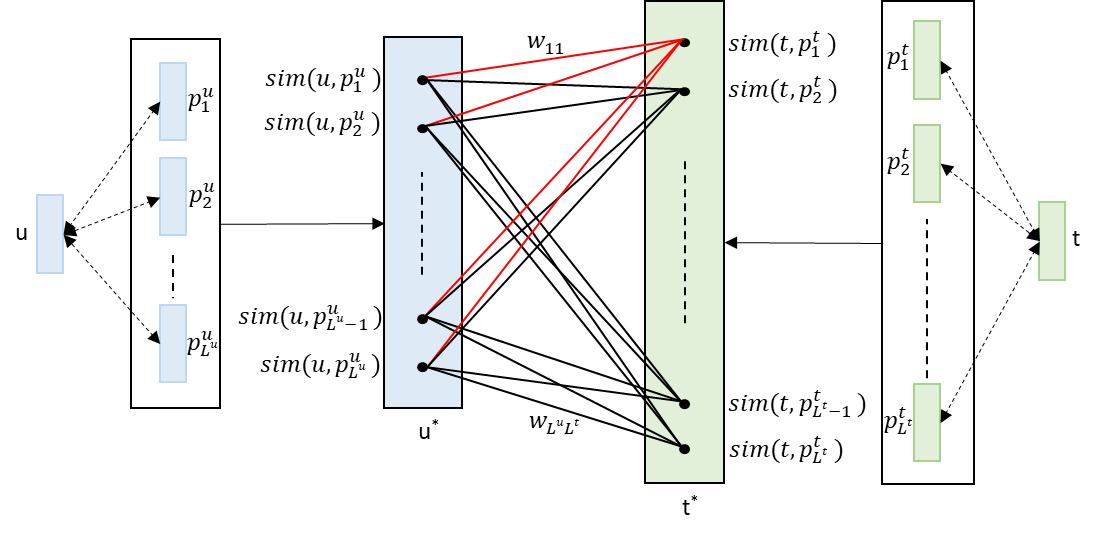}
  \caption{UIPC-MF Model. The red lines show the used connection weights to obtain user preference value for a particular item prototype 1, $r_{1}$.}
  \label{figure:3.1}
\end{figure}

Finally, UIPC-MF utilizes connection weights to link ${u^{*}}$ and ${t^{*}}$ in order to calculate the final predicted logit for $(u,t)$. The connection weights assign a learnable weight $w\in\mathbb{R}$ to each pair $(sim(u,p^{u}),sim(t,p^{t}))$. This implies that UIPC-MF assumes each user prototype has some degree of relation with each item prototype. For example, the user prototype representing a metal lover may exhibit a strong association with the item prototype of the metal genre and a weaker association with the classical genre. It is important to note that the connection weights are global and shared by all users and items. The overall architecture of UIPC-MF is depicted in Figure \ref{figure:3.1}. 
The formula for generating the predicted logit of $(u,t)$ is shown in equation \ref{eq:3.2}.

\begin{equation} \label{eq:3.2}
UIPC\text{-}Score(u,t) = \sum_{i=1}^{L^{u}}\sum_{j=1}^{L^{t}}w_{ij}sim(u,p_{i}^{u})sim(t,p_{j}^{t})
\end{equation}
where $w_{ij}$ represents the connection weight between the user prototype $p^{u}_{i}$ and item prototype $p^{t}_{j}$.

Equation \ref{eq:3.2} demonstrates that the predicted logit is a linear combination of $sim(u,p^{u})$ and $sim(t,p^{t})$. For each user $u$, ${u^{*}}$ and the connection weights can be utilized to calculate the user's preference value $r$ over an item prototype $p^{t}_j$ according to equation \ref{eq:3.3}.

\begin{equation} \label{eq:3.3}
r_{j} = \sum_{i=1}^{L^{u}}w_{ij}sim(u,p_{i}^{u})
\end{equation}
where, $r_{j}$ represents the preference value of users toward an item prototype $p^{t}_{j}$. It is computed by summing the similarities between a user and all user prototypes, multiplied by the corresponding connection weights, for a specific item prototype. This calculation provides a clear insight into the user preference toward the corresponding item prototype. The red lines in Figure \ref{figure:3.1} indicate the connection weights of users against the first item prototype as $r_{1}$. A higher absolute value of $r_{j}$ indicates a stronger association between user and item prototype $p^{t}_{j}$. Consequently, the user preference values for all item prototypes can be represented as $\{r_{1}, ..., r_{L^{t}}\}$. 

\begin{equation} \label{eq:3.4}
UIPC\text{-}Score(u,t) = \sum_{j=1}^{L^{t}}s_{j}, \; s_{j}=r_{j} \cdot sim(t,p_{j}^{t})
\end{equation}

In Equation \ref{eq:3.4}, $s_{j}$ denotes the final score of item prototype $j$ for the user u. The predicted logit can also be represented as the sum of all $s_{j}$. This equation provides a clear explanation of how the UIPC-Score is composed and computed as well as the roles of user preference and item prototype similarity in calculating the score.

\subsection{Loss function} 
To train the UIPC-MF model, we adopt the loss function $\mathcal{L}_{total}(\mathcal{I},\Theta)$ as equation \ref{eq:3.5}, given the data $\mathcal{I}$ with the model parameters $\Theta$.
\begin{equation} \label{eq:3.5}
\begin{split}
\mathcal{L}_{total}(\mathcal{I},\Theta)=\mathcal{L}_{base}(\mathcal{I},\Theta)+\lambda_{L2} \left\|\Theta \right\|_{2}+\lambda_{1}R_{\{\mathcal{P}^{u}\to \mathcal{U}\}}+\\
\lambda_{2}R_{\{\mathcal{U}\to \mathcal{P}^{u}\}}+\lambda_{3}R_{\{\mathcal{P}^{t}\to \mathcal{T}\}}+\lambda_{4}R_{\{\mathcal{T}\to \mathcal{P}^{t}\}}+\lambda_{L1} \left\|r \right\|_{1}
\end{split}
\end{equation}
where $\mathcal{L}_{base}(\mathcal{I},\Theta)$ can be binary cross-entropy loss (BCE Loss), Bayesian personalized ranking loss (BPR Loss), or sampled softmax loss (SSM Loss), is commonly used in implicit feedback for recommender systems \cite{Rendle21, Wu22}. We describe the details of these losses below. $\Theta$ refers to the model parameters, and $\left\|\Theta \right\|$ indicates the $L^{2}$-norm using the hyperparameter $\lambda_{L2}$ as a regularization term. The following four terms are additional interpretability terms from \cite{Melchiorre22}, and the last term is the $L^{1}$-norm for users' preference scores. 

During training, for each observed interaction $(u,t)$, we sample a set as negative examples, denoted by $\mathcal{N}=\{n_{1},n_{2},...,n_{\left|\mathcal{N} \right|}\}$. We use $I$ and $I^{-}$ to denote the all observed interactions and the all sampled negative interactions, respectively. For each user $u$, $\mathcal{I}_{u}=\{i_{1},i_{2},...,i_{\left|\mathcal{I}_{u} \right|}\}$ denotes the item set that the user $u$ has interacted with. The function $f$ denotes the UIPC-MF which outputs the logit score, and $\sigma$ denotes the logistic sigmoid function $\sigma(x)=\frac{1}{1+e^{-x}}$.
\subsubsection{Binary Cross-Entropy Loss}

The binary cross-entropy loss is defined as follows:
\begin{equation} \label{eq:3.6}
\begin{split}
\mathcal{L}_{BCE}=-\sum_{(u,t)\in I}^{}\log\sigma (f(u,t))-\\
\sum_{(u,n)\in I^{-}}^{}\log(1-\sigma (f(u,n)))
\end{split}
\end{equation}
This is a pointwise loss, which means that it treats the recommendation problem as a binary classification or regression problem. \cite{Rendle21, Wu22} The BCE loss measures the distance between the predicted and true probability distributions.
\subsubsection{Bayesian Personalized Ranking Loss}

The Bayesian personalized ranking loss \cite{Rendle09} is defined as follows:
\begin{equation} \label{eq:3.7}
\mathcal{L}_{BPR}=-\frac{1}{\left|I \right|}\sum_{u=1}^{N}\sum_{t\in \mathcal{I}_{u},n\notin \mathcal{I}_{u}}^{} \log\sigma (f(u,t)-f(u,n))
\end{equation}
This is a pairwise loss, which means that it treats the recommendation problem as a ranking task. The BPR loss maximizes the score difference between observed items and unobserved items for each user. Specifically, it aims to learn a ranking such that observed items for each user are ranked higher than unobserved items. \cite{Rendle21, Wu22}
\subsubsection{Sampled Softmax Loss}

The sampled softmax loss is defined as follows:
\begin{equation} \label{eq:3.8}
\mathcal{L}_{SSM}=-\frac{1}{\left|I \right|}\sum_{(u,t)\in I}^{}\log\frac{\exp(f(u,t))}{\exp(f(u,t))+\sum\limits_{n\in \mathcal{N}}^{}\exp(f(u,n))}
\end{equation}
The SSM loss is used as a substitute for the softmax loss by considering only a small subset of items within the loss. Its purpose is to assign a relatively higher probability to observed interactions compared to sampled negative interactions in the softmax function \cite{Rendle21, Wu22}.
\subsubsection{Interpretability Terms}

The following four terms are interpretability terms utilized in ProtoMF \cite{Melchiorre22}. Their definitions are as follows:
\begin{equation} \label{eq:3.9}
R_{\{\mathcal{P}^{u}\to \mathcal{U}\}} = -\frac{1}{L^{u}}\sum_{l=1}^{L^{u}}\max_{i\in [1,...,N]}sim(u_{i},p_{l}^{u})
\end{equation}
\begin{equation} \label{eq:3.10}
R_{\{\mathcal{U}\to \mathcal{P}^{u}\}} = -\frac{1}{N}\sum_{i=1}^{N}\max_{l\in [1,...,L^{u}]}sim(u_{i},p_{l}^{u})
\end{equation}
\begin{equation} \label{eq:3.11}
R_{\{\mathcal{P}^{t}\to \mathcal{T}\}} = -\frac{1}{L^{t}}\sum_{l=1}^{L^{t}}\max_{j\in [1,...,M]}sim(t_{j},p_{l}^{t})
\end{equation}
\begin{equation} \label{eq:3.12}
R_{\{\mathcal{T}\to \mathcal{P}^{t}\}} = -\frac{1}{M}\sum_{j=1}^{M}\max_{l\in [1,...,L^{t}]}sim(t_{j},p_{l}^{t})
\end{equation}

The first term, $R_{\{\mathcal{P}^{u}\to \mathcal{U}\}}$, ensures that each user prototype is correlated with at least one user embedding. This is accomplished by maximizing the similarity between each user prototype and the user with the highest similarity value. The second term, $R_{\{\mathcal{U}\to \mathcal{P}^{u}\}}$, ensures that each user embedding is correlated with at least one user prototype. This is achieved by maximizing the similarity between each user and the prototype with the highest similarity value. The concept of the third term, $R_{\{\mathcal{P}^{t}\to \mathcal{T}\}}$, is the same as the first term, guaranteeing that each item prototype is correlated with at least one item embedding. Similarly, the last term, $R_{\{\mathcal{T}\to \mathcal{P}^{t}\}}$, ensures that each item embedding is correlated with at least one item prototype.

\subsubsection{$L^{1}$-Norm for Users' Preference Scores}
\begin{equation} \label{eq:3.13}
\left\|r \right\|_{1} = \frac{1}{N}\sum_{i=1}^{N}\sum_{j=1}^{L^{t}}\left|r_{ij} \right|
\end{equation}
The last term as shown in equation \ref{eq:3.13} is $L^{1}$-Norm for users' preference values that promotes the convergence of preference values $r$ towards zero, ensuring that each preference value for an item prototype does not have a strong bias towards either a strictly positive or negative range for all users. Instead, we aim for each user's preference value for an item prototype to be positive or negative depending on their individual preferences. This helps prevent prototypes from becoming a bias during training. 

To reduce computation time, we only calculate the users and items that appear in the sampled batch during training. Hyperparameters $\lambda_{1}$, $\lambda_{2}$, $\lambda_{3}$,  $\lambda_{4}$ and $\lambda_{L1}$ are used to adjust the influence of these  terms.

\section{Experiments and Discussion}

\begin{table}
\centering
\footnotesize
\begin{tabular}{ llll } 
    \toprule
    & ML-1M &AMAZONVID &MO-3MON \\
    \midrule
        \# Users & 6,034&6,950& 8,385   \\
        \# Items & 3,125&14,494& 27,696   \\
        \# Interactions & 574,376&132,209& 921,194   \\
    \bottomrule
\end{tabular}
\captionof{table}{Statistics of the filtered datasets for collaborative filtering}
\label{table:1}
\end{table}

Our experiments for collaborative filtering include the following datasets which cover three domains:

\textbf{MovieLens-1M (ML-1M):} \cite{Harper15}
The movie rating dataset comprises 1 million ratings on a scale from 1 to 5. We adopted the settings described in \cite{Melchiorre22}. Ratings above 3.5 are considered positive interactions, and a 5-core filtering approach was applied to both users and items, which involved filtering out users and items with fewer than 5 occurrences in the interactions.

\textbf{Amazon Video Games (AMAZONVID):} \cite{Mcauley15}
The video game rating dataset is rated on a scale from 1 to 5. We also adopted the settings described in \cite{Melchiorre22}. Ratings above 3.5 are considered positive interactions, and a 5-core filtering approach was applied to both users and items. 

\textbf{Music4All-Onion-3Months (MO-3MON):} \cite{Moscati22} This dataset is derived from the LFM-2b dataset\cite{Schedl22}, which is a large dataset of music listening events. We utilized the listening events recorded during the last three months, specifically from January 1, 2020, to March 19, 2020. Following the filtering approach LFM2B-1MON described in \cite{Melchiorre22}, we retained users whose age ranges from 10 to 95 and for whom gender information is available. Outlier users who listened to more than the 99th percentile compared to all users were removed. Finally, a 10-core filtering approach was applied to both users and items.

The statistical information of the filtered datasets is presented in Table \ref{table:1}.

\subsection{Evaluation Metrics} \label{sec4.3}
\subsubsection{Hit Ratio (HR)}
The Hit Ratio measures the proportion of true interaction pairs that rank within a specified cutoff number among other negatively sampled interactions with true interaction pairs in the ranked list. In our experiments, the ranked list corresponds to either the validation data or the test data.
The formula of HR@K is below:
\begin{equation} \label{eq:4.1}
HR@K = \frac{\left|I^{K}_{hit} \right|}{\left|I_{all} \right|}
\end{equation}
where $\left|I^{K}_{hit} \right|$ represents the number of true interaction pairs ranked within the top K positions in the ranked list, and $\left|I_{all} \right|$ represents the total number of ranked items in the ranked list. \cite{Benjamin21}

\subsubsection{Normalized Discounted Cumulative Gain (NDCG)}
 NDCG is a widely used evaluation metric in recommender systems that measures the quality of a ranked list of items based on their relevance to a user's preferences. The NDCG score ranges from 0 to 1, where a higher score indicates a better ranking quality. If the true interaction pairs (items relevant to the user) are ranked closer to the top of the list, the NDCG score will be higher.

\subsection{Baseline Models} \label{sec4.2}
We compared the performance of three baseline methods: \textit{Matrix Factorization (MF)} \cite{Koren09,Rendle09}, a classical collaborative filtering method, and two prototype-related methods \textit{Anchor-based Collaborative Filtering}\cite{Barkan21} (ACF) and \textit{Prototype-based Matrix Factorization}\cite{Melchiorre22} (ProtoMF).

\subsection{Training Details} \label{sec4.4}
In this section, we describe the training details. First, we employ the leave-one-out strategy to obtain the train/validation/test data. This means we retrieve the second from last and the last interacted items sorted by timestamp as validation and test data, respectively, for each user. For evaluation, we sample 99 negative interactions for each validation and test interaction. The number of negative interactions sampled for each interaction in the training set is a hyperparameter.

To optimize all models, we use the mini-batch optimizers: Adam \cite{kingma14} and Adagrad \cite{Duchi11}.

\subsubsection{Hyperparameter Tuning}


We followed the hyperparameter tuning method and value range described in \cite{Melchiorre22} for all models in our experiments. All hyperparameter types and ranges in UIPC-MF are the same as ProtoMF \cite{Melchiorre22}, except for $\lambda_{L1}$ which has the same range as $\lambda_{L2}$. We used Tree-structured Parzen Estimators \cite{Bergstra11,Bergstra13} to tune each model by sampling 100 hyperparameter configurations. The maximum number of training epochs was set to 100. To save training time, we utilized the HyperBand trial-scheduler \cite{Li17} for both the ML-1M and MO-3MON datasets. Training was stopped if the HR@10 metric on the validation set did not improve for 10 consecutive epochs. The model with the highest HR@10 was selected as the best model. We repeated this process three times using different seeds and reported the average performance against the test set.

\subsection{Evaluation Results} \label{sec4.5}

\begin{table*}[ht]
\centering
\scriptsize
\centering
\begin{tabular}{@{\extracolsep{4pt}}lcccccccccccc}
    \toprule   
     {} & \multicolumn{4}{c}{ML-1M}  & \multicolumn{4}{c}{AMAZONVID}& \multicolumn{4}{c}{MO-3MON}\\ 
     \cmidrule{2-5} 
     \cmidrule{6-9} 
     \cmidrule{10-13}
      Model  & HR@5 & HR@10 & N@5 & N@10 & HR@5 & HR@10 & N@5 & N@10 & HR@5& HR@10 & N@5 & N@10\\ 
    \midrule
    MF  & 0.344 & 0.501 & 0.230 & 0.280 & 0.149 & 0.234 & 0.102 & 0.130 & 0.221& 0.324 & 0.149 & 0.182\\ 
    ACF  & 0.412 & 0.596 & 0.275 & 0.334 & 0.221 & 0.360 & 0.138 & 0.183 & 0.477& 0.653 & 0.330 & 0.387\\ 
    ProtoMF  & 0.458 & 0.640 & 0.308 & 0.367 & 0.269 & 0.404 & 0.179 & 0.223 & 0.511& 0.670 & 0.364 & 0.416\\ 
    UIPC-MF  & 0.476 & 0.656 & 0.325 & 0.383 & 0.274 & 0.411 & 0.181 & 0.225 & 0.549& 0.702 & 0.398 & 0.447\\ 
    UIPC-MF-$L^{1}$  & \textbf{0.490} & \textbf{0.669} & \textbf{0.335} & \textbf{0.394} & \textbf{0.283} & \textbf{0.419} & \textbf{0.187} & \textbf{0.231} & \textbf{0.558}& \textbf{0.710} & \textbf{0.405} & \textbf{0.455}\\
    \bottomrule
\end{tabular}
\captionof{table}{Recommendation evaluation results}
\label{table:2}
\end{table*}

To empirically investigate \textbf{RQ1}, we evaluated UIPC-MF with and without the $L^{1}$-Norm. Table \ref{table:2} demonstrates that UIPC-MF outperforms all baseline models on the three datasets, and UIPC-MF-$L^{1}$ further improves performance across all datasets. In-depth analysis of the results show that ACF outperforms MF on all datasets, and ProtoMF outperforms ACF. Compared with the best-performing model ProtoMF in the baselines, UIPC-MF-$L^{1}$ improves on average by 4.7\% in HR@10, 6.8\% in NDCG@10, 7.1\% in HR@5 and 8.2\% in NDCG@5 across the three datasets. UIPC-MF-$L^{1}$ significantly improves performance against all other models, especially in the cutoff 5, which means it has the ability to rank true interactions at the top in comparison to other models.

The tables \ref{table:6}, \ref{table:7} and \ref{table:8} in Appendix display the optimal hyperparameters and their corresponding test results for UIPC-MF-$L^{1}$.

\subsection{Explaining UIPC-MF Recommendations}
\begin{figure*}[h]

  \centering
  \includegraphics[width=\linewidth]{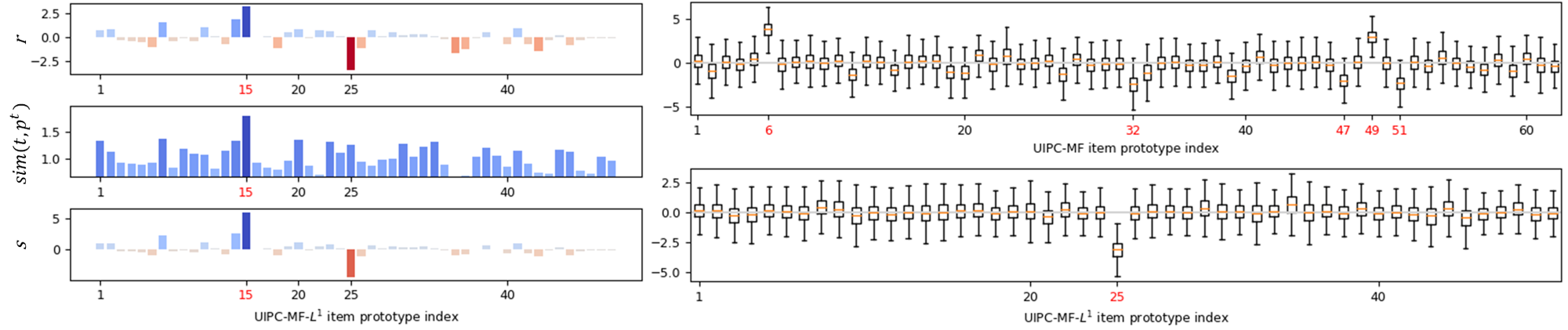}
  \caption{The left side shows the visualization of the user preference value $r$, $\text{sim}(t,p^t)$, and the item prototype score $s$, while the right side presents the boxplots with the range of user preference values in UIPC-MF (top) and UIPC-MF-$L^1$ (bottom).}
  \label{figure:2}
\end{figure*}

To answer \textbf{RQ2}, we visualize an example of MO-3MON on Figure \ref{figure:2}, which corresponds to Equation \ref{eq:3.4}. In this example, the user has the highest preference for item prototype 15, and the item has the highest similarity to prototype 15. As a result, the score of item prototype 15 accounts for the majority of the weight in the predicted score. 

Since the item prototypes and items are in the same space, we can calculate the top nearest items with the item prototype. The left column of Table \ref{table:4} shows the top 10 nearest items to item prototype 15, including their song names, genres, and total occurrences in the dataset. All of them close to the metal genre. 

In this example, where the user has the highest preference for item prototype 15, we can obtain the top songs with the highest similarity to prototype 15 that the user has listened to in the training dataset. Table \ref{table:3} shows the top 10 nearest items to item prototype 15 that the user has listened to in the training dataset.

This information can be used as the explanation rationales for the recommendation. For example:\textit{"Since you love metal music, you may also enjoy the music we are recommending."} or \textit{"Other listeners who have listened to XXX also enjoyed the music we are recommending."} This example demonstrates the transparency of the proposed model, as it allows the recommender systems to access the underlying inference of the model in generating a recommendation item for a specific user.

\begin{table}
\centering
\footnotesize
\begin{tabular}{ cc } 
    \toprule
    Song& Genre \\
    \midrule
        \textbf{Where the Slime Live} & \textit{death metal}   \\
        \textbf{Grotesque Impalement} & \textit{death metal}   \\
        \textbf{Severed Survival} & \textit{death metal}   \\
        \textbf{Far Below} & \textit{progressive rock}   \\
        \textbf{You Can't Bring Me Down} & \textit{thrash metal}   \\
        \textbf{Them Bones} & \textit{grunge}   \\
        \textbf{Hell Awaits} & \textit{thrash metal}   \\
        \textbf{The Burning Shadows of Silence} & \textit{symphonic black metal}   \\
        \textbf{Silent Scream} & \textit{thrash metal}   \\
        \textbf{Jesus Built My Hotrod} & \textit{industrial metal}   \\
    \bottomrule
\end{tabular}
\captionof{table}{The top 10 nearest items to item prototype 15 that the user has listened to in the training dataset.}
\label{table:3}
\end{table}

\begin{table}
\centering
\footnotesize
\begin{tabular}{ c|c } 
    \toprule
    Item prototype 15&Item prototype 25 \\
    \midrule
        \multicolumn{1}{l|}{\textbf{Mask of the Red Death}} &  \multicolumn{1}{l}{\textbf{Low Grade Buzz}}   \\
        \multicolumn{1}{r|}{ \textit{black metal/15}} & \multicolumn{1}{r}{\textit{indietronica/11}}   \\
        
        \multicolumn{1}{l|}{\textbf{Unhallowed}} &  \multicolumn{1}{l}{\textbf{Airsick}}   \\
        \multicolumn{1}{r|}{ \textit{black metal/36}} & \multicolumn{1}{r}{\textit{ambient/11}}   \\
        
        \multicolumn{1}{l|}{\textbf{Crown}} &  \multicolumn{1}{l}{\textbf{Betrayed in the Octagon}}   \\
        \multicolumn{1}{r|}{ \textit{black metal/15}} & \multicolumn{1}{r}{\textit{ambient/10}}   \\
        
        \multicolumn{1}{l|}{\textbf{Grim and Frostbitten Kingdoms}} &  \multicolumn{1}{l}{\textbf{Hymn to Eternal Frost}}   \\
        \multicolumn{1}{r|}{ \textit{black metal/25}} & \multicolumn{1}{r}{\textit{black metal/9}}   \\
        
        \multicolumn{1}{l|}{\textbf{Release From Agony}} &  \multicolumn{1}{l}{\textbf{Blue Drive}}   \\
        \multicolumn{1}{r|}{ \textit{thrash metal/13}} & \multicolumn{1}{r}{\textit{ambient/12}}   \\

        \multicolumn{1}{l|}{\textbf{Heaving Earth}} &  \multicolumn{1}{l}{\textbf{Criminals}}   \\
        \multicolumn{1}{r|}{ \textit{death metal/20}} & \multicolumn{1}{r}{\textit{indie rock/12}}   \\

        \multicolumn{1}{l|}{\textbf{Sworn to the Black}} &  \multicolumn{1}{l}{\textbf{Baby}}   \\
        \multicolumn{1}{r|}{ \textit{death metal/21}} & \multicolumn{1}{r}{\textit{folk/9}}   \\  

        \multicolumn{1}{l|}{\textbf{Certain Death,Sadus}} &  \multicolumn{1}{l}{\textbf{A Lesson Never Learned}}   \\
        \multicolumn{1}{r|}{ \textit{thrash metal/17}} & \multicolumn{1}{r}{\textit{metalcore/12}}   \\ 

        \multicolumn{1}{l|}{\textbf{Flag of Hate,Kreator}} &  \multicolumn{1}{l}{\textbf{Applesauce}}   \\
        \multicolumn{1}{r|}{ \textit{thrash metal/36}} & \multicolumn{1}{r}{\textit{indie rock/8}}   \\ 

        \multicolumn{1}{l|}{\textbf{The Day You Died}} &  \multicolumn{1}{l}{\textbf{Vanishing Point}}   \\
        \multicolumn{1}{r|}{ \textit{death metal/20}} & \multicolumn{1}{r}{\textit{dark ambient/11}}   \\         
    \bottomrule
\end{tabular}
\captionof{table}{The top 10 nearest items in item prototypes 15 and 25}
\label{table:4}
\end{table}

\subsection{The impact of $L^{1}$-Norm in reduction of learning bias}

To assess the impact of $L^{1}$-Norm on the model and answer \textbf{RQ3}, we present boxplots of users' preferences for item prototypes in the MO-3MON dataset. Figure \ref{figure:2} shows that UIPC-MF-$L^{1}$ exhibits less bias towards extreme positive or negative values in comparison to UIPC-MF. All users in UIPC-MF exhibit strictly positive or negative preference values towards item prototypes $6$, $32$, $47$, $49$, and $51$ within a specified range. However, prototype $25$ in UIPC-MF-$L^{1}$ still has negative preferences for all users. We list the top 10 nearest items to item prototype $25$ in the right column of Table \ref{table:4}. The occurrences are approximately equal to the preprocessing filtering threshold (10), and their genres lack significant consistency. This indicates a negative bias towards long tail items learned by UIPC-MF-$L^{1}$. This phenomenon aligns with common observations in collaborative filtering \cite{Abdollahpouri19, Sankar21}, and it is observed within the internal workings of UIPC-MF-$L^{1}$.

\subsection{Parameter Complexity}
We compare the parameter complexities of the baseline models against the UIPC-MF model, as discussed in \cite{Melchiorre22}. We assume a simple setting for all models, including $N$ users and $M$ items. Additionally, ACF has $K$ anchors, while ProtoMF and UIPC-MF have $K$ user prototypes and $K$ item prototypes. We assume each user embedding vector, item embedding vector, anchor vector, and prototype vector is a $d$-dimensional vector. Then the MF model contains $(N + M) \times d$ parameters for $N$ users and $M$ items. ACF adds $K \times d$ parameters for the $K$ anchors. ProtoMF has an additional $2K \times d$ parameters for the user and item prototypes, as well as extra $2K \times d$ parameters for the linear transformation weights. UIPC-MF has an additional $2K \times d$ parameters for the user and item prototypes, along with $K^2$ parameters for the connection weights. Since $N$ and $M$ are much larger than $K$ and $d$, the additional parameters in ACF, ProtoMF, and UIPC-MF constitute only a small portion compared to the parameters required for $N$ users and $M$ items. We display the parameter count and the ratio of additional parameter count to MF parameter count (APC/MF ratio) for each model with $N=5000$, $M=5000$, $K=100$, and $d=100$ in Table \ref{table:5}.

\begin{table}
\centering
  \small
  \begin{tabular}{lll}
    \toprule
    Model&Parameter count&APC/MF ratio\\
    \midrule
    MF&1,000,000&0\\
    ACF&1,010,000&0.01\\
    ProtoMF&1,040,000&0.04\\
    UIPC-MF&1,030,000&0.03\\   
  \bottomrule
\end{tabular}
\captionof{table}{Parameter count in each model}
\label{table:5}
\end{table}

From the Table \ref{table:5}, it shows that the additional parameters in ACF, ProtoMF, and UIPC-MF constitute only a small portion compared to the MF parameters.

\section{Conclusion and Future work}

\subsection{Conclusion}

This paper presents a novel prototype-based collaborative filtering that offers several contributions. First, the proposed approach learns both user and item representations from the same data space as the corresponding prototypes. Second, the predicted logit score is determined by a linear combination of the crossover interactions between user prototype similarities and item prototype similarities. This allows our model to reveal the preference of each user against each item prototype and the association relationships between user prototypes and item prototypes using the connection weights. Third, to enhance performance and alleviate bias, we employ the $L^{1}$-Norm on user preference to avoid bias towards positive or negative preference values for all users. Fourth, the transparency of the model reveals a negative bias towards long tail items for all users. Finally, through experimental results on three datasets, we demonstrate the superiority of our model in comparison to other prototype-based methods in top-N recommendation tasks, establishing it as a state-of-the-art solution.

\subsection{Future work}
For future research directions, we can utilize the prototype in explainable content-based recommendation. Content-based recommendation enhances representations by incorporating more information such as user and item features, as well as content data of items. Perhaps the prototype used in content-based recommendation can help us better understand the effects of this information on recommendations. Additionally, since the prototype can learn general attributes from the training data, we can combine it with content-based recommendation to address the cold start problem in recommender systems. For example, we can use the prototype to learn the general attributes from the music audio signal. When a new music item is added to the music recommender system, the system can compute an estimated score for all users based on the prototypes of the music audio signal.

Since the experiments are conducted using implicit feedback, it is also possible to apply UIPC-MF to explicit feedback. By leveraging the transparency of UIPC-MF, we may be able to uncover the reasons behind users' preferences or dislikes for certain items, which differ from the reasons why users may interact with items in this paper.

\bibliography{aaai24}

\appendix
\section{Optimal Hyperparameters for UIPC-MF-$L^{1}$ } \label{apB}
The tables below display the optimal hyperparameters with their corresponding test results for UIPC-MF-$L^{1}$. Since we repeated the tuning process three times using different seeds for each model, there are three sets of hyperparameter combinations for each dataset.

\begin{center}
    \scriptsize
    \begin{tabular}{lllll}
    \toprule
    &&\#1&\#2&\#3\\
    \midrule
    \multicolumn{1}{c}{\multirow{15}{*}{Hyperparameters}}& Embedding size & 33&49&90\\
    & $\lambda_{L2}$ & .134173e-2&.951028e-3&.336879e-3\\
    & Base loss & SSM&SSM&SSM\\
    & Sampling & uniform&uniform&uniform\\
    & Neg. samples & 37&48&28\\
    & Batch size & 124&142&126\\
    & Optimizer & Adagrad&Adagrad&Adam\\
    & LR & .226839e-1&.168761e-1&.373224e-3\\
    & \# User prototypes & 84&59&66\\
    & \# Item prototypes & 95&88&69\\
    & $\lambda_{1}$ & .153054e-1&.592179&.849859\\
    & $\lambda_{2}$ & .134081e-1&.104536e-2&.365722e-1\\
    & $\lambda_{3}$ & .188699&.240550e-1&.755034e-1\\
    & $\lambda_{4}$ & 1.322822&.341394&.712384e-1\\
    & $\lambda_{L1}$ & .318446e-2&.271684e-2&.534349e-2\\
    \hline
    \multicolumn{1}{c}{\multirow{4}{*}{Results}}& NDCG@5 & .3409&.3372&.3276\\
    & NDCG@10 & .3983&.3978&.386\\
    & HR@5 & .4988&.4914&.4783\\
    & HR@10 & .6757&.6725&.6588\\

  \bottomrule
\end{tabular}
\captionof{table}{Optimal Hyperparameters for UIPC-MF-$L^{1}$ in ML-1M}
\label{table:6}
\end{center}

\begin{center}
    \scriptsize
    \begin{tabular}{lllll}
    \toprule
    &&\#1&\#2&\#3\\
    \midrule
    \multicolumn{1}{c}{\multirow{15}{*}{Hyperparameters}}& Embedding size & 93&33&84\\
    & $\lambda_{L2}$ & .385810e-3&.157954e-3&.380064e-2\\
    & Base loss & BCE&BCE&BCE\\
    & Sampling & popular&popular&popular\\
    & Neg. samples & 45&46&16\\
    & Batch size & 75&382&71\\
    & Optimizer & Adagrad&Adam&Adagrad\\
    & LR & .404478e-1&.157421e-2&.406145e-1\\
    & \# User prototypes & 85&38&54\\
    & \# Item prototypes & 73&36&28\\
    & $\lambda_{1}$ & .456862e-2&1.306350&5.926725\\
    & $\lambda_{2}$ & .336176e-2&.141989e-2&.902861e-2\\
    & $\lambda_{3}$ & .191007&.274330e-2&.111797e-2\\
    & $\lambda_{4}$ & .473007&.137768&.749137e-1\\
    & $\lambda_{L1}$ & .204884e-1&.161991e-1&.273646e-3\\
    \hline
    \multicolumn{1}{c}{\multirow{4}{*}{Results}}& NDCG@5 & .1958&.1814&.1834\\
    & NDCG@10 & .2395&.227&.2253\\
    & HR@5 & .2934&.2777&.2788\\
    & HR@10 & .4285&.4191&.4086\\

  \bottomrule
\end{tabular}
\captionof{table}{Optimal Hyperparameters for UIPC-MF-$L^{1}$ in AMAZONVID}
\label{table:7}
\end{center}

\begin{center}
    \scriptsize
    \begin{tabular}{lllll}
    \toprule
    &&\#1&\#2&\#3\\
    \midrule
    \multicolumn{1}{c}{\multirow{15}{*}{Hyperparameters}}& Embedding size & 56&51&55\\
    & $\lambda_{L2}$ & .175263e-2&.365098e-3&.190545e-3\\
    & Base loss & SSM&SSM&BPR\\
    & Sampling & uniform&uniform&uniform\\
    & Neg. samples & 19&47&40\\
    & Batch size & 146&130&148\\
    & Optimizer & Adagrad&Adagrad&Adam\\
    & LR & .163911e-1&.698419e-1&.123853e-3\\
    & \# User prototypes & 99&41&71\\
    & \# Item prototypes & 50&96&77\\
    & $\lambda_{1}$ & .171384e-2&.302235e-1&.563671e-1\\
    & $\lambda_{2}$ & .436756e-2&.125097e-1&.350694\\
    & $\lambda_{3}$ & .761652e-2&.290637&.147196\\
    & $\lambda_{4}$ & .686550e-2&.449296e-2&1.143978\\
    & $\lambda_{L1}$ & .139693e-2&.732059e-3&.704644e-3\\
    \hline
    \multicolumn{1}{c}{\multirow{4}{*}{Results}}& NDCG@5 & .4135&.4155&.3873\\
    & NDCG@10 & .4614&.463&.4401\\
    & HR@5 & .5671&.5666&.5403\\
    & HR@10 & .7134&.7123&.7035\\

  \bottomrule
\end{tabular}
\captionof{table}{Optimal Hyperparameters for UIPC-MF-$L^{1}$ in MO-3MON}
\label{table:8}
\end{center}
\end{document}